\documentclass[aps,prd,nofootinbib]{revtex4}
\usepackage[utf8]{inputenc}
\usepackage[english]{babel}
\usepackage[T1]{fontenc}

\usepackage{amsmath}
\usepackage{amsfonts}
\usepackage{amssymb}
\usepackage{graphicx}

\usepackage{lmodern}

\usepackage{hyperref}
\hypersetup{colorlinks, 
citecolor=blue,
linkcolor=red,
urlcolor=black}

\begin{document}

\title{Quantum Brownian motion for a particle in analog expanding cosmologies in the presence of disclination}
\author{E. J. B. Ferreira}
\email{ejbf@academico.ufpb.br}
\author{E. R. Bezerra de Mello}
\email{emello@fisica.ufpb.br}
\author{H. F. Santana Mota}
\email{hmota@fisica.ufpb.br}
\affiliation{Departamento de Física, Universidade Federal da Paraíba, Caixa Postal 5008, João Pessoa, Paraíba, Brazil}


\begin{abstract}
In this paper we study the quantum brownian motion of a scalar point particle in the analog Friedman-Robertson-Walker spacetime in the presence of a disclination, in a condensed matter system. The analog spacetime is obtained as an effective description of a Bose-Einstein condensate in terms of quantum excitations of sound waves, named phonons. The dynamics of the phonons is described by a massless real scalar field whose modes are also subjected to a quasi-periodic condition. In this sense, we find exact solutions for the real scalar field in this scenario and calculate the two-point function which makes possible to analyze the mean squared velocity dispersion of the particle in all directions. We, thus, analyze some interesting particular cases and show some graphs where it is possible to see the consistency of our results. 
\end{abstract}


\maketitle

\section{Introduction}

Quantum brownian motion (QBM) of a point particle is a topic that has been investigated in several contexts and approaches \cite{gour1999will,deLorenci2014quantum,camargo2018vacuum, yu2004vacuum,yu2004brownian,bessa2009brownian}. This is an example of a phenomenon category induced by modifications in the quantum vacuum fluctuations of a field, along with the famous Casimir effect \cite{bordag2009advances,dalvit2011casimir}. In the QBM the quantum fluctuations of the vacuum state of a field, for instance, the scalar \cite{gour1999will,deLorenci2014quantum,camargo2018vacuum} and electromagnetic \cite{yu2004vacuum,yu2004brownian,bessa2009brownian} ones, induce a stochastic motion on a particle. The latter, in turn, behaves like a probe device of the quantum vacuum fluctuation effects, which reveal themselves when modifications in the Minkowski free vacuum comes about as a consequence of the imposition of boundary conditions, thermal effects, and so on. In order to make the studies increasingly realistic and explore the effects causing the QBM of the particle over time, different elements have been investigated such as temperature corrections \cite{deLorenci2019, yu2006brownian, deLorenci2021probing}, smoothing interactions through switching functions \cite{camargo2018vacuum, camargo2019vacuum}, which regularize typical divergences in the model, and the wave packet structure of the particle \cite{seriu2009smearing}. It is important to point out that although both QBM and Classical brownian motion are two types of stochastic motion, they have distinct characteristics due to their quantum and classical origins, respectively. One of the main diferences, for example, is that in the quantum domain negative dispersions are possible to happen, a fact that in the literature is known as subvacuum effect (see \cite{deLorenci2021probing} and references therein).

General Relativity as well as Quantum Field Theory predict highly sensitive physical phenomena, for example, gravitational waves and Hawking radiation. The experimental verification of such phenomena requires highly advanced and accurate technology, which is certainly the case of gravitational waves. Although gravitational waves have been predicted a long time ago, only recently  they were detected experimentally \cite{abbott2016observation}. On the other hand, Hawking radiation has been observed only in gravity analog models \cite{steinhauer2016observation}, since in gravitational context it faces serious difficulties due to its small magnitude when compared to the cosmic radiation background.

Gravity analog models have also been the subject of studies in different aspects and scenarios. At a more elementary level, for instance, those analog systems are constituted by fluids (as water) in which the acoustic perturbation velocity, that is, the sound wave, in the medium, plays a role analogous to the light velocity in a spacetime geometry. In other words, sound waves propagating through a fluid behave like the light in curved spacetime. In addiction, it is possible to make this analogy in more complex systems such as Bose-Einstein condensates (BEC) \cite{barcelo2011analogue}.

BECs are quantum systems widely used in the literature to investigate various types of phenomena. In fact, for instance, in Ref. \cite{eckel2018rapidly} the radial expansion of a ring-shaped BEC has been considered to simulate a real cosmological expanding universe, and among the investigations carried out, phonons redshift is studied, which consists in the analogous effect to photons in real cosmological spacetime. We may also quote the study of the particle production in a BEC in different expanding scenarios \cite{jain2007}. The BECs implementation in gravitational wave detections is an idea that has been discussed \cite{robbins2019bose}. So with these examples we can see that BECs are indeed very interesting systems to be considered.

The QBM of point particles also can be studied in the gravity analog model scenarios previously mentioned. For instance, in Ref. \cite{bessa2017quantum} it was studied the QBM of a point particle in the effective spacetime like the one of the Friedman-Robertson-Walker (FRW) spacetime in Cosmology, in the absence of boundary, as well as in the presence of plane boundaries. In summary, the authors studied two distinct cases: free particle and bounded particle by a classical nonfluctuating external force that cancel the expanding effects. A remarkable result is that in the bounded particle case, in the absence of plane boundaries, a constant and isotropic velocity dispersion is obtained, indicating the existence of a stochastic motion. Here, we decided to study a similar system, but now considering the presence of a disclination without plane boundaries, so that we can be able to analyze different configurations from the ones in Ref. \cite{bessa2017quantum}.

More specifically, in the present paper, we propose to study the QBM of a massless scalar point particle in an analog model simulating a conformal expanding universe in the presence of a linear topological defect known in condensed matter as disclination, and also requiring that the scalar field obeys a quasiperiodic condition on the angular variable. In order to simulate the expanding universe we consider an expanding BEC system that has been explored as a mean of investigation for different physical phenomena. In this sense, we could envisage a condensed matter system, like liquid crystals, where a disclination is possible to exist. We should mention that disclination is the analog, in condensed matter, to cosmic string \cite{satiro2005liquid}, arising due to phase transitions in the early Universe. Cosmic strings as it is known in the literature may have several cosmological, astrophysical and gravitational implications, observationally \cite{copeland2011seeking,hindmarsh2011signals,mota2015big}.

Regarding to the structure of this paper, in Section \ref{sec02} we present a brief discussion about the main relations that allow us to establish the effective expanding spacetime for the phonons present in the BEC, and taken into consideration the presence of a disclination. In Section \ref{sec03} we provide the massless scalar field solution, obeying a quasiperiodic condition on the angular variable, in the effective FRW spacetime considered. In Section \ref{sec04} we obtain the positive frequency two-point Wightman function, which is an essential element  for our computation of the mean squared velocity dispersion (MSVD) of the particle. The point particle velocity equation coupled to the massless scalar field is given in Section \ref{sec05}, which we use to obtain the velocity dispersion in Section \ref{sec06} and to study its behavior. Finally, in Section \ref{sec07conclusion}, we summarize the main results in this paper and present our Conclusions.

\section{Effective FRW spacetime in Bose-Einstein condensate with a disclination}\label{sec02}

The condensation occurs when particles of a boson gas (e.g. rubidium atoms) are submitted to extremely low temperatures so that they tend to agglomerate in the lowest energy state of the system, usually called ground state. 
The exact description of this system is made through the so called second quantization formalism, where the fields associated to the system are quantized and become operators. Thereby,
it is known that the field operator $\hat{\Psi}(\mathbf{x},t)$ related to the bosons condensate obeys the equation of motion \cite{jain2007, barcelo2001, dalfovo1999}
\begin{eqnarray}\label{eq101}
		i\hbar\dfrac{\partial\hat{\Psi}(\mathbf{x},t)}{\partial t} = \left[-\dfrac{\hbar^{2}}{2m}\nabla^{2}+V_{\mathrm{ext}}(\mathbf{x}) + U |\hat{\Psi}(\mathbf{x},t)|^{2} \right]\hat{\Psi}(\mathbf{x},t),
	\end{eqnarray}
where $V_{\mathrm{ext}}$ represents an external potential which affect equally all particles in the gas, for example, the particle confinement potential. The term $U=4\pi\hbar^{2}a/m$ is a two-body potential interaction, i.e. between two bosons in the condensate, which is written in terms of the bosons mass and the scattering length $a$. It is important to mention that the parameter $a$ will play a key role in the mathematical construction below, because through it will be defined a velocity for the acoustics perturbations on the condensate.
	
Although the exact description of the dynamics of the system involves operators, as shown in Eq. \eqref{eq101}, if the gas is weakly interacting and lies in the low temperatures regime, the dominante occupation by the particles in a single state (the ground state) suggest that the field $\hat{\Psi}(\mathbf{x},t)$ allows the decomposition \cite{jain2007}
\begin{eqnarray}\label{eq102}
\hat{\Psi}(\mathbf{x},t) = \Psi(\mathbf{x},t) +\delta\hat{\varphi}(\mathbf{x},t),
\end{eqnarray}
where $\Psi(\mathbf{x},t)$ correspond to the mean field value, $\Psi(\mathbf{x},t)=\langle\hat{\Psi}(\mathbf{x},t)\rangle$.
The point to be noted is that since all particles lies in the ground state, we can describe the system on average by a collective wave function. It is because of this collective description that sometimes $\Psi(\mathbf{x},t)$ is said to be the wave function of the condensate \cite{barcelo2001}. In this approach $\delta\hat{\varphi}(\mathbf{x},t)$ represent the quantum part of the boson field operator, $\hat{\Psi}$, which is interpreted as quantum or thermal fluctuations with null mean value, that is, $\langle\delta\hat{\varphi}(\mathbf{x},t)\rangle=0$. This mean-field approach is known as Bogolyubov approximation \cite{jain2007}.

Using Eqs. \eqref{eq101} and \eqref{eq102}, and neglecting the fluctuations, we can establish that
\begin{eqnarray}\label{eq103}
                i\hbar\dfrac{\partial\Psi(\mathbf{x},t)}{\partial t} = \left[-\dfrac{\hbar^{2}}{2m}\nabla^{2}+V_{\mathrm{ext}}(\mathbf{x}) + U \left\vert\Psi(\mathbf{x},t) \right\vert^{2}  \right]\Psi(\mathbf{x},t),
        \end{eqnarray}
which is called Gross-Pitaevskii equation. This equation will be the basis to obtain a mathematical description of the analog model taken into account in our investigation.
Note that unlike Eq. \eqref{eq101}, which involves operators, Eq. \eqref{eq103} is written in terms of the classical field, which corresponds to the wave function describing the atoms in the ground state of the condensate.
Furthermore, it is instructive to note the similarity between \eqref{eq103} and the Schrödinger equation for a particle of mass $m$ submitted to a potential $V_{\mathrm{ext}}$, but with a nonlinear term that is responsible for the two-body interaction.

\subsection{Effective metric}

From the classical point of view the starting point to obtain an effective metric are both the continuity and Euler equations \cite{barcelo2011analogue}. Similarly, in the quantum level, we can get a effective metric by considering in Eq. \eqref{eq103} the field decomposition \cite{jain2007}
\begin{eqnarray}\label{eq104}
\Psi(\mathbf{x},t)=\sqrt{n(\mathbf{x},t)}e^{i\theta(\mathbf{x},t)},
\end{eqnarray}
where $n(\mathbf{x},t)$ and $\theta(\mathbf{x},t)$ are the density and the phase classical real fields, respectively. One should remember that in general the wave function is a complex object, and complex quantities can be written in terms of modulus and phase (the polar form of a complex number).

Substituting Eq. \eqref{eq104} in Eq. \eqref{eq103}, after some algebraic manipulations, it is possible to show that smalls perturbations in the phase field of the condensate wave function experiences an apparently curved spacetime geometry. This analog spacetime is characterized by the effective metric \cite{jain2007}
\begin{eqnarray}\label{eq109c}
g_{\mu\nu}(\mathbf{x},t)=\left(\dfrac{n_{0}}{cm}\right)^{\frac{2}{(d-1)}}\left[
	\begin{array}{ccc}
		-(c^{2}-v^{2}) & \vdots &-v_{j}\\
		\cdots & & \cdots\\
		-v_{i}&\vdots &\delta_{ij}
	\end{array}\right],
\end{eqnarray}
where $d$ represents spatial dimensions,
\begin{eqnarray}
\mathbf{v}=\dfrac{\hbar}{m}\nabla\theta_{0}
\end{eqnarray}
is a background velocity and $c$ is the speed of the acoustic perturbations in the condensate, given by 
\begin{eqnarray}\label{eq109b}
c^{2}=\dfrac{Un_{0}}{m}.
\end{eqnarray}
The parameters $\theta_{0}$ and $n_{0}$ are background values of the phase and density fields, respectively. We would like to emphasize that the metric \eqref{eq109c} is established in the so called acoustic approximation, also known as hydrodynamics approximation or semiclassical approximation. In this regime, the quantum pressure term is negligible which leads to an effective metric that describes an analog Lorentzian geometry \cite{jain2007} (see also Ref. \cite{barcelo2001} for other approximations). Note that in the regime of the acoustic approximation the geometry plays a central role and typical parameters, like the healing length, are absent.

\subsection{Effective expanding spacetime with a disclination}

To create an effective geometry that simulates an expanding spacetime from the effective metric \eqref{eq109c}, we follow distinct routes \cite{barcelo2003analogue}. In particular, one of these is done by varying the acoustic perturbation velocity in the condensate, by considering a temporal dependence in the sound velocity propagation. This method is  commonly used in the literature \cite{jain2007, bessa2017quantum, anacleto2021stochastic,barcelo2003probing} and here we will adopted it\footnote{For an example of a distinct method see Ref. \cite{weinfurtner2005analogue}.}.

Similar to Refs. \cite{jain2007,bessa2017quantum, anacleto2021stochastic}, here the time variation in the sound velocity will be introduced by means of a time dependence of the scattering length, $a=a(t)$, and the atoms density $n_{0}$ is maintained fixed in the condensate. Hence, Eq. \eqref{eq109b} becomes
\begin{eqnarray}
c(t)^{2} = \dfrac{4\pi\hbar^{2}n_{0}}{m}a(t).
\label{ctime}
\end{eqnarray}
Now we want to establish that the time variation takes place through a dimensionless scale function responsible for modulating the interaction about some reference value. In order for this to happen we can rewrite Eq. \eqref{ctime} as
\begin{eqnarray}
c(t)^{2} = \left(\dfrac{4\pi\hbar^{2}n_{0}}{m}a_{0}\right)b(t) = U_{0}b(t).
\end{eqnarray}
Thus, the relation between the sound velocity and the scaling function $b(t)$ is 
\begin{eqnarray}
b(t) = \left[\dfrac{c(t)}{c_{0}}\right]^{2}.
\end{eqnarray}
Here, $U_{0}=U(t=t_0)$, $a_{0}=a(t=t_0)$ and $c_{0}=c(t=t_0)$ correspond to constant values at the initial time $t_{0}$ and, consequently $b(t_{0})=1$. It is important to emphasize that, in this approach, a simulating expanding spacetime is related to the capacity to control the velocity of the acoustic perturbations.

Considering the preceding discussions, from the metric \eqref{eq109c} with $v=0$ \footnote{As indicated in Ref. \cite{weinfurtner2008phenomenon} although a time variation is admitted in sound velocity the condensate stay at rest so the background velocity is zero.}, we get the four-dimensional line element \cite{jain2007}
\begin{eqnarray}\label{eq110a}
		d\overline{s}^{2}_{\textrm{eff}}=\Omega_{0}^{2}\left[-c_{0}^{2}b^{\frac{1}{2}}(t)dt^{2}+b^{-\frac{1}{2}}(t)\delta_{ij}dx^{i}dx^{j}\right],
	\end{eqnarray}
with
	\begin{eqnarray}
		\Omega_{0}^{2} &=& \dfrac{n_{0}}{c_{0}m}.
	\end{eqnarray}
As indicated in Ref. \cite{bessa2017quantum} we note that due to the dimensional constant $\Omega_{0}^{2}$, this line element has different dimension from that of a square length, but we can regain the usual dimension redefining the line element. Since $\Omega_{0}^{2}$ is a constant factor we redefine the line element \eqref{eq110a} such that $d\overline{s}^{2}_{\textrm{eff}}\Omega_{0}^{-2}=ds^{2}_{\textrm{eff}}$. Therefore
\begin{eqnarray}\label{eq110b}
ds^{2}_{\textrm{eff}}=-c_{0}^{2}b(t)^{\frac{1}{2}}dt^{2}+b^{-\frac{1}{2}}(t)\delta_{ij}dx^{i}dx^{j}.
\label{lineeff}
\end{eqnarray}
Furthermore, this line element allows us to make the conformal time transformation 
\begin{eqnarray}\label{eq110c}
dt = b^{-\frac{1}{2}}(t)d\eta .
\end{eqnarray} 
As it is our interest to consider a system with a disclination, the suitable symmetry for the spatial part of the line element \eqref{lineeff} is the cylindrical one, that is,
\begin{eqnarray}\label{eq110d}
\delta_{ij}dx^{i}dx^{j} = d\rho^{2}+\rho^{2}d\phi +dz^{2},
\end{eqnarray}
where $\rho \geq 0$, $\phi \in [0,2\pi/p]$ and $z\in (-\infty,\infty)$. Here the parameter $p\neq 1$ characterizes the presence of the disclination in the medium, or in the analog model context the existence of the cosmic string\footnote{Diferently from the cosmic string, the parameter $p$ associated with disclination can also  assumes values smaller than unity.}. Note that it modifies the range of the angular variable, $\phi$, in contrast to the usual cylindrical coordinates where $p=1$. From Eqs. \eqref{eq110b}, \eqref{eq110c} and \eqref{eq110d} we get
\begin{eqnarray}\label{eq110e}
ds^{2}_{\textrm{eff}}=b^{-\frac{1}{2}}(\eta)[-c_{0}d\eta^{2}+d\rho^{2}+\rho^{2}d\phi +dz^{2}].
\end{eqnarray} 
This line element characterizes an effective FRW spacetime in presence of a disclination (cosmic string analog) which is conformally related to the locally flat effective disclination spacetime, with conformal factor $b^{-\frac{1}{2}}(\eta)$.

\section{Klein-Gordon equation and normalized solutions}\label{sec03}

The Klein-Gordon (KG) equation for the massless scalar field $\psi(x)$ in curved spacetime, non-minimally coupled to gravity, is given by
\begin{eqnarray}\label{eq401}
\dfrac{1}{\sqrt{-g}}\partial_{\mu}\sqrt{-g}g^{\mu\nu}\partial_{\nu}\psi + \xi R\psi =0,
\end{eqnarray}
where $R$ is the Ricci scalar and $\xi$ the nonminimal curvature coupling. Two specific values of $\xi$ are: $\xi=0$ and $\xi=1/6$, that correspond to the minimal and conformal ccouplings in $(1+3)-$dimensional spacetime \cite{birrell1984quantum}, respectively. Note also that, in the context of the condensate discussed in the previous section, $\psi(x)$ will represent the quantum fluctuating parte of the phase $\theta(x)$ introduced in Eq. \eqref{eq104}  \cite{jain2007}. Therefore, $\psi$ is a real massless scalar field that describes the phonons excitations in the condensate. It is important to stress that Eq. \eqref{eq401} is not the standard form that describes the perturbations in the field phase $\theta(x)\equiv\psi(x)$ of the BEC. In fact, the equation which arises from the Gross-Pitaevskii equation \eqref{eq103} by implementing the acoustic approximation is given by Eq. \eqref{eq401} without the term $\xi R\psi$ (nonminimal coupling to gravity). In any case, we are assuming the existence of such a coupling. Nonetheless, it is possible that this nonminimal coupling can be introduced via the generalized Gross-Pitaevskii equation shown in Ref. \cite{barcelo2001}.  Another possibility is through some modification in the external potencial in \eqref{eq103}.

In the case of conformal transformations the metric tensor and scalar field obey the following relations,  
\begin{eqnarray}\label{eq402a}
\tilde{g}_{\mu\nu}(x)=\Omega^{2}(x)g_{\mu\nu}(x)
\end{eqnarray}
and
\begin{eqnarray}\label{eq402b}
\tilde{\psi}(x)=\Omega^{\frac{(2-n)}{2}}(x)\psi(x)  \   ,
\end{eqnarray}
being $\Omega(x)$ a positive, real and nonvanishing function and $n$ is the dimension number of the spacetime \cite{birrell1984quantum}. For our case, $n=4$ and $\Omega^{2} \equiv b^{-\frac{1}{2}}$. 

For massless bosonic field and conformally coupled to the geometry, the two-point function in an expanding spacetime in the presence of disclination can be obtained by the corresponding one in Mikowiski spacetime. In order to construct the two-point function, the equations \eqref{eq402a} and \eqref{eq402b} allow us to solve the KG equation in the disclination spacetime, described by the line element in square brackets in Eq. \eqref{eq110e}, and then to obtain the solution in the expanding spacetime with a disclination, by making use of a conformal transformation.

The positive frequency solution of the KG equation in the presence of a disclination spacetime is given by 
\begin{eqnarray}\label{eq403}
\psi (t,\rho ,\phi ,z) =Ae^{-i\omega t+ipk_{\phi}\phi+ik_{z}z}J_{|k_{\phi}|}(l\rho)  \  ,
\end{eqnarray}
where $A$ is a normalization constant and $J_{\mu}(z)$ the Bessel function \cite{gradshtein2007}. The energy is given by the expression
\begin{eqnarray}
\omega=\sqrt{c_{0}^{2}l^{2}+k_{z}^{2}} \ .
\end{eqnarray}

Here we want to consider that the massless scalar field obeys the quasiperiodic condition in the angular variable,
\begin{eqnarray}\label{eq404}
\psi(t,\rho,\phi,z) = e^{-i(2\pi\beta) }\psi(t,\rho,\phi +2\pi/p,z),
\end{eqnarray}
where $0 \leq \beta <1$ stands for the quasiperiodicity of the solution. Let us now point out that the condition above, as discussed in Ref. \cite{kretzschmar1965must}, indicates in our case that the acoustic field is a multi-valued function as a consequence of the parameter $\beta$ being a noninteger number. As also argued in Ref. \cite{kretzschmar1965must}, there exists an equivalence of continuous multi-valued functions with discontinues single-valued ones. This makes possible to establish a connection with the famous Aharonov-Bohm effect. Since the parameter $\beta$ is the cause of the discontinuity it may be interpreted as an interaction \cite{de2012topological}  of an external source with a single-valued acoustic field. In the case of the original Aharonov-Bohm effect the interaction is between a line of magnetic flux with an electrically charged particle \cite{Aharonov:1959fk, Griffiths1995}.  In our case, on the other hand, the analog Aharonov-Bohm effect may represent scattering of phonons as a consequence of the interaction of a vortex line with the acoustic field \cite{coutant2015acoustic,sonin2010aharonov}.

Now by imposing the condition \eqref{eq404} on the solution \eqref{eq403}, we get
\begin{eqnarray}\label{eq405}
\psi(t,\rho,\phi,z) = A e^{-i\omega t+ip(m+\beta)\phi+ik_{z}z}J_{p|m+\beta |}(l\rho),
\end{eqnarray}
where $m$ is an integer number and $k_{\phi} = q(m+\beta)$. The solution above is characterized by the complete set of
quantum numbers  $\sigma = (l, m, k_z)$.

The normalization constant $A$ can be obtained by making using of the normalization condition 
\begin{eqnarray}
-\dfrac{i}{\hbar c_{0}^{2}}\int d^{d}x\sqrt{-g}\{\psi_{\sigma}(x)[\partial_{t}\psi_{\sigma '}^{*}(x)] -[\partial_{t}\psi_{\sigma}(x)]\psi_{\sigma '}^{*}(x)\} = \delta_{\sigma\sigma '}  \  ,  \label{solution}
\end{eqnarray}
where the delta symbol in the right hand side of the above equation is understood as Dirac delta function for the continuous quantum number $(k_z, \ l)$, and Kronecker delta for discrete ones, $(m)$. From \eqref{solution}, one has
\begin{eqnarray}\label{eq406}
\psi_{\sigma}(x) = \left(\dfrac{pl\hbar c_{0}^{2}}{8\pi^{2}\omega}\right)^{\frac{1}{2}} e^{-i\omega t+ip(m+\beta)+ik_{z}}J_{p|m+\beta |}(l\rho).
\end{eqnarray}
We would like to emphasize that Eq. \eqref{eq406} is a normalized solution for the KG equation in the disclination spacetime, satisfying the quasiperiodic condition \eqref{eq404}. This quasiperiodic condition allow us to obtain a generalized solution in terms of the general phase $2\pi\beta $, which takes into account other values of $\beta$ instead those of the periodic ($\beta = 0$) and antiperiodic ($\beta = 1/2$) conditions. 

Before ending this section we should point out that, in order to perform the analysis of the induced quantum brownian motion by the system under consideration, the massless real scalar field must be promoted to a field operator, $\hat{\psi}$, that can be expressed by,
\begin{eqnarray}\label{psioperator}
\hat{\psi}(x) = \sum_{\sigma}\left[a_{\sigma}\psi_{\sigma}(x) + a_{\sigma}^{\dagger}\psi_{\sigma}^{*}(x)\right],
\end{eqnarray}
where $\psi_{\sigma}(x)$ and its complex conjugate $\psi^{*}_\sigma(x)$ correspond to the complete set of normalized solutions of the KG equation, and are given by \eqref{eq406} in our case. The coefficients $a$ and $a^{\dagger}$ are, respectively, the creation and annihilation bosonic operators, that obey the commutation relation $[a_{\sigma}, a_{\sigma'}^{\dagger}] = \delta_{\sigma,\sigma'}$. The sum symbol over the set of quantum numbers $\sigma$ in Eq. \eqref{psioperator}, is defined as
\begin{eqnarray}
\sum_{\sigma} = \sum_{m=-\infty}^{\infty}\int_{-\infty}^{\infty}dk_{z}\int_{0}^{\infty}dl.
\label{sumdef}
\end{eqnarray}
Therefore, we are able now to introduce and calculate the two-point Wightman function in the next section.

\section{Wightman function}\label{sec04}

At the quantum level, one of the most important quantities in the study of quantum vacuum fluctuation effects is the positive frequency  two-point Wightman function, $\mathcal{W}$, which is defined as \cite{mota2018scalar}
\begin{eqnarray}\label{Wfun}
	\mathcal{W}(x,x') &=&  \langle \hat{\psi}(x)\hat{\psi}(x')\rangle \nonumber\\
	&=& \sum_{\sigma}\psi_{\sigma}(x)\psi_{\sigma}^{*}(x'),
\end{eqnarray}
where $\langle \ldots \rangle\equiv \langle 0|\ldots |0\rangle$ stands for the vacuum expectation value. The mean value of the fields product above is taken by making use of Eq. \eqref{psioperator}, and the commutation relation for the creation and annihilation operators. As usual the vacuum state $|0\rangle$,  is defined by the condition $a|0\rangle = 0$. This, ultimately, results in the sum over the quantum numbers $\sigma$ of the product of the field $\psi_{\sigma}(x)$ by its complex conjugate $\psi^{*}(x')$.

Substituting the normalized solution \eqref{eq406} in the definition of the Wightman function \eqref{Wfun}, along with \eqref{sumdef}, after performing the integrals in $k_z$ and $l$, provides the expression
\begin{eqnarray}\label{eq501a}
\mathcal{W}_{\text{d}} = \dfrac{\hbar pc_{0}}{8\pi^{2}\rho\rho'}e^{ip\beta\Delta\phi}\int_{0}^{\infty}d\xi e^{-\frac{\delta}{2\rho\rho '}\xi}\mathcal{I}(p,\beta , \xi)  \  ,
\end{eqnarray}
where we have defined $\delta =\Delta\tau^2 + \Delta z^2 + \rho^{2} + \rho'^{2} $, with $\Delta\tau = ic_0\Delta t$ indicating that a Wick rotation has been performed and $\Delta t = t-t'$, $\Delta z=z-z'$. In addition, the function $\mathcal{I}(p,\beta , \xi)$ is defined as 
\begin{eqnarray}\label{eq501b}
\mathcal{I}(\beta , p, \xi) = \sum_{m=-\infty}^{\infty}e^{ipm\Delta\phi}I_{p|m+\beta |}(\xi),
\end{eqnarray}
where $\Delta\phi = \phi - \phi'$. Let us point out that we have been able to perform the integral in $k_z$ and $l$ by making use of the identity 
\begin{eqnarray}\nonumber
\dfrac{e^{-\omega\Delta\tau}}{\omega} = \dfrac{2}{\sqrt{\pi}}\int_{0}^{\infty}dze^{-\omega^{2}z^{2}-\frac{\Delta\tau^{2}}{4z^{2}}}.
\end{eqnarray}
In particular, the integral in $l$ has been worked out by using the relation 
\begin{eqnarray}\nonumber
\int_{0}^{\infty}e^{-s^2z^2}J_{r}(\alpha z)J_{r}(\beta z)zdz = \dfrac{1}{2s^2}e^{-\frac{(\alpha^2+\beta^2)}{4s^2}}I_{r}\left(\dfrac{\alpha\beta}{2s^2}\right),
\end{eqnarray}
where $I_{\mu}(z)$ is the modified Bessel functions of the first kind \cite{gradshtein2007}.

The sum in Eq. \eqref{eq501b} can be performed by using the following integral representation \cite{de2015vacuum, Farias2021}:
\begin{eqnarray}\label{eq502}
\mathcal{I}(\beta , p, \xi) &=& \sum_{m=-\infty}^{\infty}e^{ipm\Delta\phi}I_{p|m+\beta |}(\xi) \nonumber\\
&=&\dfrac{1}{p}\sum_{m}e^{\xi\cos\left(\frac{2m\pi}{p}-\Delta\phi\right)}e^{i\beta (2m\pi-p\Delta\phi)}\nonumber\\
&-&\dfrac{1}{2\pi i}\sum_{j=-1}^{1}je^{j(i\pi p\beta)}\int_{0}^{\infty}dy \dfrac{\{\cosh[py(1-\beta)]-\cosh(p\beta y)e^{-ip(\Delta\phi +j\pi)}\}}{e^{\xi\cosh(y)}\{\cosh(py)-\cos[p(\Delta\phi +j\pi)]\}},
\end{eqnarray}
where the $m$ index summation in the right hand side goes under the restriction 
\begin{eqnarray}\nonumber
-\dfrac{p}{2}+\dfrac{\Delta\phi}{(2\pi/p)}\leq m \leq\dfrac{p}{2}+\dfrac{\Delta\phi}{(2\pi/p)}.
\end{eqnarray}
Moreover, from Eqs. \eqref{eq501a} and \eqref{eq502} we obtain
\begin{eqnarray}\label{eq503}
\mathcal{W}_{\mathrm{d}}(x,x')&=& \dfrac{\hbar c_{0}}{4\pi^{2}}\sum_{m}\dfrac{1}{\sigma_{m}} e^{2\pi m\beta i}\nonumber\\
&-&\dfrac{\hbar c_{0}p}{8\pi^{3}i}\int_{0}^{\infty}dy\dfrac{1}{\sigma_{y}}\mathcal{F}(\beta ,p,\Delta\phi,y),
\end{eqnarray}
where
\begin{eqnarray}\label{eq504a}
\sigma_{m}&=&\delta -2\rho\rho'\cos\left(\dfrac{2m\pi}{p}-\Delta\phi\right),\nonumber\\
\sigma_{y}&=&\delta +2\rho\rho'\cosh(y),
\end{eqnarray}
and
\begin{eqnarray}\label{eq504b}
\mathcal{F}(\beta ,p,\Delta\phi,y) &=& \sum_{j=-1}^{1}je^{j(i\pi p\beta)}\dfrac{\{\cosh[py(1-\beta)]-\cosh(p\beta y)e^{-ip(\Delta\phi +j\pi)}\}}{\{\cosh(py)-\cos[p(\Delta\phi +j\pi)]\}}.
\end{eqnarray}
Having obtained the positive frequency Wightman function, we are in position to calculate the MSVD of the particle for each component.

\section{Equation of Motion}\label{sec05}

In order to obtain the equation of motion for a point particle in the FRW spacetime in the presence of a disclination, let us consider the expression, in curved spacetime,
\begin{eqnarray}\label{eq201a}
m\dfrac{Du^{\mu}}{d\tau} = qg^{\mu\nu}\nabla_{\nu}\psi + f^{\mu}_{\mathrm{ext}},
\end{eqnarray}
where $u^{\mu}$ is a four vector velocity, $m$ the mass of the point particle and $q$ its charge. This is the equation of motion to a point particle coupled to a massless scalar field $\psi$ \cite{mota2020induced, bessa2017quantum}. Here backreaction effects have been neglected so we could solve a homogenous KG equation. Furthermore,
\begin{eqnarray}\label{eq201b}
\dfrac{Du^{\mu}}{d\tau} =\dfrac{du^{\mu}}{d\tau}+\Gamma^{\mu}_{\alpha\beta}u^{\alpha}u^{\beta}
\end{eqnarray}
is the covariant derivative of the four vector  $u^{\mu}$ and
\begin{eqnarray}\label{eq201c}
\Gamma^{\mu}_{\alpha\beta} = \dfrac{1}{2}g^{\gamma\mu}\left(g_{\gamma\alpha ,\beta}+g_{\gamma\beta ,\alpha}-g_{\alpha\beta ,\gamma}\right),
\end{eqnarray}
are the Christoffel symbol.

Let us now consider the equation of motion \eqref{eq201a} in the nonrelativistic regime, which is perfectly reasonable for the study of the QBM. In this case, the proper time $\tau$ and the time coordinate $t$ are practically the same and only the spatial components of Eq. \eqref{eq201a} are significant. Hence, Eq. \eqref{eq201a} becomes
\begin{eqnarray}\label{eq202}
m\dfrac{Du^{i}}{dt} = f^{i} + f^{i}_{\mathrm{ext}},
\end{eqnarray}
where $f^{i}=qg^{ij}\partial_{j}\psi$ is the force arising due to the existence of the field $\psi$. The term $f^{i}_{\mathrm{ext}}$ accounts for other possible external force contributions to the dynamics of the charged scalar particle of mass $m$. These contributions are in general of classical origin, e.g., the gravitational and electromagnetic forces. Note that, in essence, Eq. \eqref{eq202} is very similar to the Langevin equation for the classical brownian motion of a suspended particle in a fluid. Classically, $f^{i}_{\mathrm{ext}}$ is associated with nonfluctuating forces while $f^{i}$ with a stochastic force. However, here, in the context of quantum field theory, the force $f^{i}$ carries the information of the quantum vacuum fluctuations of the scalar field $\psi$.

Taking into consideration the line element \eqref{eq110e}, the only nonzero components of the Christoffel symbols are
\begin{eqnarray}\label{eq203}
\Gamma^{\rho}_{0\rho}&=&\Gamma^{\rho}_{\rho 0}=\Gamma^{\phi}_{0\phi}=\Gamma^{\phi}_{\phi 0}=\Gamma^{z}_{0z}=\Gamma^{z}_{0 z} = -\dfrac{1}{4}\dfrac{\dot{b}}{b},\\
\Gamma^{\rho}_{\phi\phi}&=&-\rho , \ \Gamma^{\phi}_{\rho\phi}=\Gamma^{\phi}_{\phi\rho}=\dfrac{1}{\rho},
\end{eqnarray}
where $\dot{b}$ stands for the time derivate of the function $b(t)$.

Since $f^{i}_{\mathrm{ext}}$ accounts for classical force contributions, in order to focus only on quantum effects on the particle motion, it is suitable to consider
\begin{eqnarray}\label{eq204}
f^{i}_{\mathrm{ext}}=m(\Gamma^{\rho}_{\phi\phi}u^{\phi}u^{\phi},2\Gamma^{\phi}_{\rho\phi}u^{\rho}u^{\phi},0).
\end{eqnarray}
Note that if we make the identification $u^{\rho}=v$ and $u^{\phi}=\omega =v/\rho$ we realize that $f^{\rho}_{\mathrm{ext}}$ and $f^{\phi}_{\mathrm{ext}}$ are both similars to centripetal forces, which are classical. Based on this, it is plausible to consider the form of $f^{i}_{\mathrm{ext}}$ presented in Eq. \eqref{eq204}, so that we can be able to focus only on quantum fluctuation effects, arising from $f^{i}$.

Carry on with our analysis, from Eqs. \eqref{eq202}, \eqref{eq203} and \eqref{eq204}, we obtain 
\begin{eqnarray}\label{eq205}
u^{i}(t)=\dfrac{qb^{\frac{1}{2}}(t)}{m}\int_{t_0}^{t}dtb^{-\frac{1}{2}}(t)g^{ij}\partial_{j}\psi,
\end{eqnarray}
where we have considered a zero value for the initial velocity, that is, $u^{i}(t_{0}) = 0$. Thus, Eq. \eqref{eq205} provides the coordinate velocity expression to a point particle in an expanding spacetime with a disclination. Note that the information about the spacetime under consideration is codified in the metric tensor present in \eqref{eq205}.
%
\section{Velocity Dispersion}\label{sec06}
\subsection{General expression}

We are now interested in calculating the velocity dispersion, that is, the MSVD of the particle. The latter, in turn, is obtained by using the definition 
\begin{eqnarray}\label{eq301_adicional}
\langle (\Delta u^{i})^{2} \rangle = \langle \left(u^{i}(x)\right)^2\rangle - \langle u^{i}(x)\rangle^2.
\end{eqnarray}
It is clear that by making the scalar field becoming an operator, the velocity of the point particle in Eq. \eqref{eq205} also becomes an operator, resulting in $\langle u^{i}(t)\rangle=0$ since $\langle\hat{\psi}(x)\rangle=0$. This is straightforward to obtain by using Eq. \eqref{psioperator} and the act of the creation and annihilation operators on the vacuum state. The velocity dispersion of the particle in Eq. \eqref{eq301_adicional} is, thus, given only by the first term on the r.h.s.\;.

Normally, the process of calculating Eq. \eqref{eq301_adicional} involves the subtraction of a divergent contribution that in general, but not always, is associated with the Minkowski spacetime contribution. In this case, we formally have
\begin{eqnarray}\label{eq301}
\langle (\Delta u^{i})^{2} \rangle_{\text{ren}} = \lim_{x'\rightarrow x} \left[\langle u^{i}(x)u^{i}(x')\rangle -    \langle u^{i}(x)u^{i}(x')\rangle_{\text{div}}\right],
\end{eqnarray}
which is the renormalized mean squared deviation of the particle velocity and the second term on the r.h.s. is the divergent contribution that comes about in the coincidence limit $x'\rightarrow x$. However, as we shall see, in our case, there will be no divergent contribution as a consequence of the choice for the function $b(t)$ in \eqref{eq305}. Our result, in fact, will be given by three contributions, where one of them is a finite constant contribution that is independent of the parameters $p$ and $\beta$. The two other contributions will, of course, depend on these parameters. Let us see below how this follows. 

As the calculation of Eq. \eqref{eq301_adicional} requires only the knowledge of the first term on the r.h.s., from Eq. \eqref{eq205}, the MSVD of the particle can be calculated through 
\begin{eqnarray}\label{eq302}
\langle u^{i}(x)u^{k}(x')\rangle  = \dfrac{q^{2}b(t)}{m^2}\int_{t_0}^{t}dt_{2}\int_{t_0}^{t}dt_{1}b^{-\frac{1}{2}}(t_{2})b^{-\frac{1}{2}}(t_{1})g_1^{ij}g_2^{kj'}\partial_{j}\partial_{j'}\mathcal{W}(x,x')_{\text{FRW}},
\end{eqnarray}
where $x=(t_1,\rho,\phi, z)$ and $x'=(t_2,\rho',\phi', z')$. The mathematical object $\mathcal{W}(x,x')_{\text{FRW}}$ correspond to the positive frequencies Wightman function, in FRW spacetime with a disclination. Moreover, we can make use of the conformal symmetry \eqref{eq402a} exhibited by the line element \eqref{eq110e}, with $\Omega^2 = b^{-\frac{1}{2}}(t)$ being the conformal factor. Thus, since we know the Wightman function \eqref{eq503} in the disclination spacetime, we can write the Wightman function in an expanding spacetime with a disclination through the conformal relation \cite{mota2020induced}
\begin{eqnarray}\label{eq303}
\mathcal{W}(x,x')_{\text{FRW}} = b^{\frac{1}{4}}(t_{1})b^{\frac{1}{4}}(t_{2})\mathcal{W}(x,x')_{\text{d}}.
\end{eqnarray}
Thereby, by applying the conformal time transformation \eqref{eq110c} in Eq. \eqref{eq302} and using \eqref{eq303}, we get
\begin{eqnarray}\label{eq304}
\langle u^{i}(x)u^{k}(x')\rangle  = \dfrac{q^{2}b(\eta)}{m^2}\int_{\eta_{0}}^{\eta}d\eta_{2}\int_{\eta_0}^{\eta}d\eta_{1} b^{-\frac{3}{4}}(\eta_{2})b^{-\frac{3}{4}}(\eta_{1})g_{1}^{ij}g_{2}^{kj'}\partial_{j}\partial_{j'}\mathcal{W}(x,x')_{\text{d}},
\end{eqnarray}
which provides the velocity dispersion, or in other words, the MSVD of the particle in an expanding spacetime in the presence of a disclination. Note that the Wightman function $\mathcal{W}(x,x')_{\text{d}}$ and the metric tensor components $g_{ij}$ are given by Eqs. \eqref{eq503} and \eqref{eq110e}, respectively.

In order to calculate Eq. \eqref{eq304} is necessary to pick a direction $i$ and an appropriate $b(\eta)$ function to simulate the expansion, i.e., to model the dynamics of the acoustic perturbation as the condensate expands. Here we will consider the asymptotically flat scaling factor
\begin{eqnarray}\label{eq305}
b^{\frac{1}{4}}(\eta)=b^{\frac{1}{4}}_{0}+b^{\frac{1}{4}}_{1}\tanh\left(\dfrac{\eta}{\tau}\right),
\end{eqnarray}
where $\tau$ is a constant time responsible for controlling the rate of expansion. Note that we must have $b^{\frac{1}{4}}_{0} > b^{\frac{1}{4}}_{1}$ in order to ensure that the line element \eqref{eq110e} is nonsingular for all real values of $\eta$. The dimensionless constants $b^{\frac{1}{4}}_{0}$ and $b^{\frac{1}{4}}_{1}$ can be defined in terms of the asymptotic values of the scaling function which determine their asymptotic bounds, e.g. the beginning and the end of the expansion, where the scaling function is a constant. We should emphasize that this is a scaling factor frequently used in the literature since in many cases it allows us to give an analytic treatment to Eq. \eqref{eq304} \cite{bessa2009brownian,jain2007,bessa2017quantum,anacleto2021stochastic}.

\subsection{MSVD in the $\rho$-direction}
%
We now wish first to calculate the $\rho$-component of the MSVD of the particle by using Eq. \eqref{eq304}.  Thus, we have
\begin{eqnarray}\label{eq306}
\langle (\Delta v^{\rho})^{2} \rangle = \dfrac{q^{2}b_{\text{f}}^{\frac{1}{2}}}{m^2}\int_{-\infty}^{\infty}d\eta_{2}b^{-\frac{1}{4}}(\eta_{2})\int_{-\infty}^{\infty}d\eta_{1}b^{-\frac{1}{4}}(\eta_{1})\partial_{\rho}\partial_{\rho'}\mathcal{W}(x,x')_{\text{d}},
\end{eqnarray}
where we have used the relation between the coordinate velocity $u^{\rho}$ and the physical velocity $v^{\rho}$, that is, $v^{\rho}=b_{\text{f}}^{-\frac{1}{4}}u^{\rho}$. The integrals above are not analytically solved for finite values of the conformal time $\eta$ in the limits of integration. However, as an approximation, we can analyze how the system behaves at a much later time so that we have been able to extend the limits of integration from $(\eta_0 , \eta)$ to $(-\infty,\infty)$. In this limit Eq. \eqref{eq305} becomes $b_{\text{f}}^{\frac{1}{4}}=b^{\frac{1}{4}}_{0}+b^{\frac{1}{4}}_{1}$, suggesting that the scalar point particle motion stopped suffering the effects of the curved spacetime. 

Hence, by substituting Eqs. \eqref{eq503} and \eqref{eq305} in \eqref{eq306}, and performing both the derivative and integration operations, in the coincidence limit $x'\rightarrow x$ we obtain\footnote{To solve the integrals we have used the residue theorem and followed an similar procedure that was done in the Appendix B of the Ref. \cite{bessa2009brownian}.},
\begin{eqnarray}\label{eq307a}
\langle (\Delta \overline{v}^{\rho})^{2} \rangle = 2\zeta(3)+2\sum_{m=1}^{\left[\frac{p}{2}\right]}\cos(2\pi m\beta)s_{m}(\chi) - \dfrac{p}{2\pi i}\int_{0}^{\infty}dy\mathcal{F}(\beta , p,y)s_{y}(\chi),
\end{eqnarray}
where $\mathcal{F}(\beta , p,y)$ is given by \eqref{eq504b} taken $\Delta\phi =0$. The result is,
\begin{eqnarray}\label{eq307b}
\mathcal{F}(\beta ,p,y) = 2i\left\{\dfrac{\cosh[py(1-\beta)]\sin(\pi p\beta)+\cosh(p\beta y)\sin[p\pi(1-\beta)]}{\cosh(py)-\cos(p\pi)} \right\},
\end{eqnarray}
and the functions $s_m(\chi)$ and $s_y(\chi)$ are given in a compact form by
\begin{eqnarray}\label{eq307d}
s_{\mu}(\chi) = 4\chi^{2}z_{\mu}^4S_{3}(\mu,\chi) - 2z_{\mu}^2S_{2}(\mu,\chi)+S_{2}(\mu,\chi) \  ,
\end{eqnarray}
with $z_{\mu}=(z_m, z_y)=\left(\sin(m\pi/p),\cosh(y/2)\right)$ and we have defined the dimensionless parameter $\chi = \frac{2\rho}{\pi c_{0}\tau}$, related to the radial distance $\rho$ to the disclination. The function $S_{\gamma}(\mu,\chi)$, on the other hand, is given by 
\begin{eqnarray}\label{eq307g}
S_{\gamma}(\mu,\chi)=\sum_{r=1}^{\infty}\dfrac{r}{[r^{2}+(\chi z_{\mu})^{2}]^{\gamma}}.
\end{eqnarray}
Note that we have also defined the dimensionless MSVD of the particle as 
\begin{eqnarray}\label{eq307c}
\langle (\Delta \overline{v}^{\rho})^{2} \rangle =\langle (\Delta v^{\rho})^{2} \rangle \left[\dfrac{2\hbar q^{2}b_{\text{f}}^{\frac{1}{2}}}{m^{2}\pi^{4}c_{0}^{3}\tau^{2}}\dfrac{\sinh^{4}(g)}{b_{1}^{\frac{1}{2}}}\right]^{-1},
\end{eqnarray}
where
\begin{eqnarray}\label{eq307f}
g = \dfrac{1}{2}\ln\left(\dfrac{\alpha^{2}+1}{\alpha^{2}-1}\right),\qquad\qquad\text{with}\qquad\qquad \alpha^{2}=\dfrac{b_{0}^{\frac{1}{4}}}{b_{1}^{\frac{1}{4}}} > 1.
\end{eqnarray}
Finally, the notation $[p/2]$ on the sum stands for the integer part of $p/2$. However, if $p$ is an integer number we should replace the sum in \eqref{eq307a} according to
\begin{eqnarray}\label{eq307j}
\sum_{m=1}^{\left[\frac{p}{2}\right]}\rightarrow \dfrac{1}{2}\sum_{m=1}^{p-1}.
\end{eqnarray}
As anticipated before, the $\rho$-component for the velocity dispersion in Eq. \eqref{eq307a} is composed by three terms, one of them being a constant (the first term in the r.h.s) that does not depend on the parameters $p$ and $\beta$, and arises due to the choice of the function $b(t)$ in Eq. \eqref{eq305}. The other two contributions carry information about the spacetime geometry and the quasiperiodicity by means of the parameters $p$ and $\beta$, respectively. Note that in Eq. \eqref{eq307a} the sum in $m$ in the second term on the r.h.s is absent if $p<2$. In the particular case that $\beta=0$ (periodicity) the two last terms on the r.h.s. of Eq. \eqref{eq307a} provide a pure disclination contribution. In contrast, if $p=1$, there will be a pure quasiperiodicity contribution given by the last term. Finally, in the case $\beta=1/2$, the last term will vanish if $p$ is an even number. This analysis will be equally valid for the other two components of the MSVD of the particle in the next sub-sections. 

\begin{figure}
\includegraphics[scale=0.5]{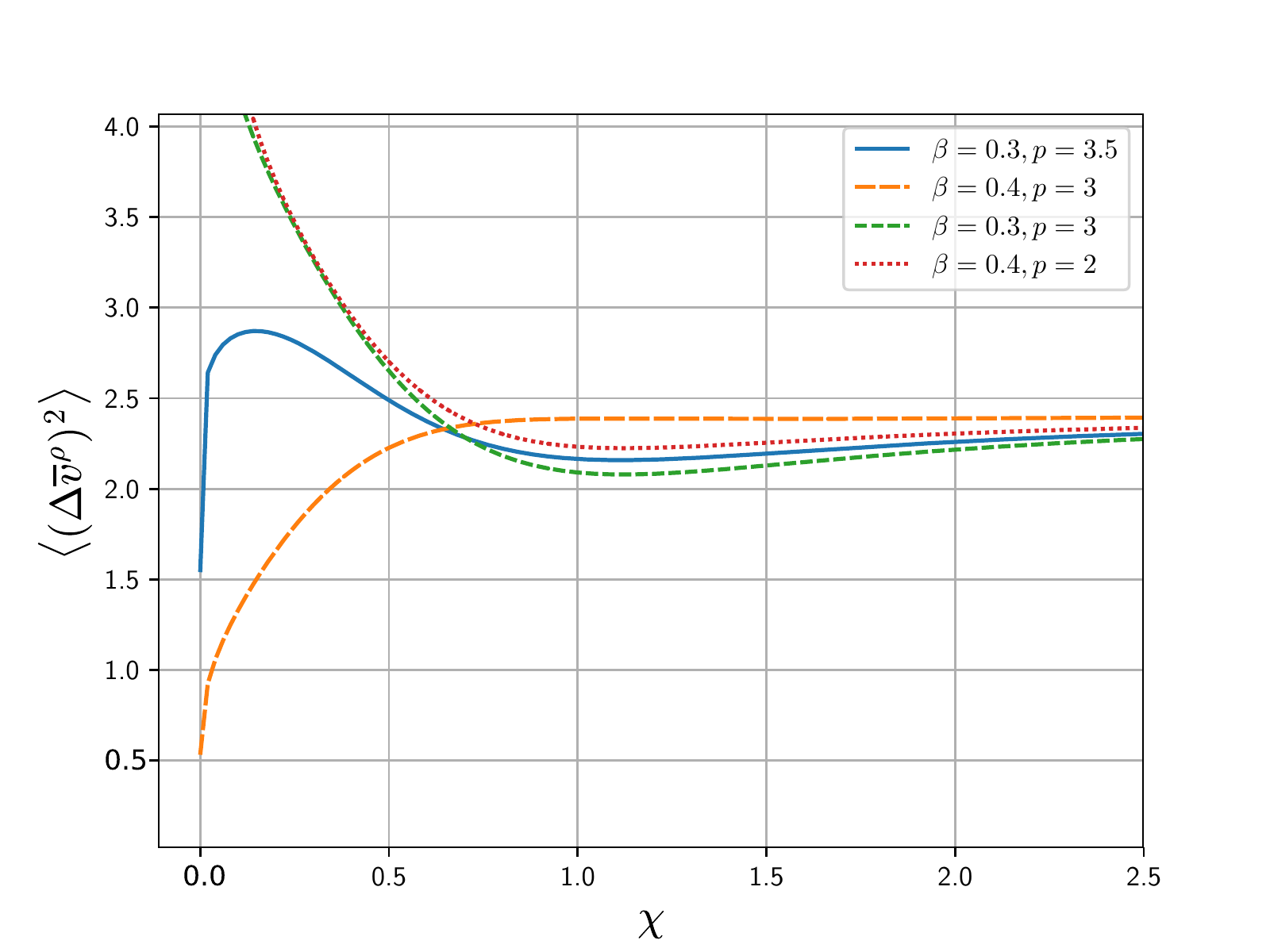}
\includegraphics[scale=0.5]{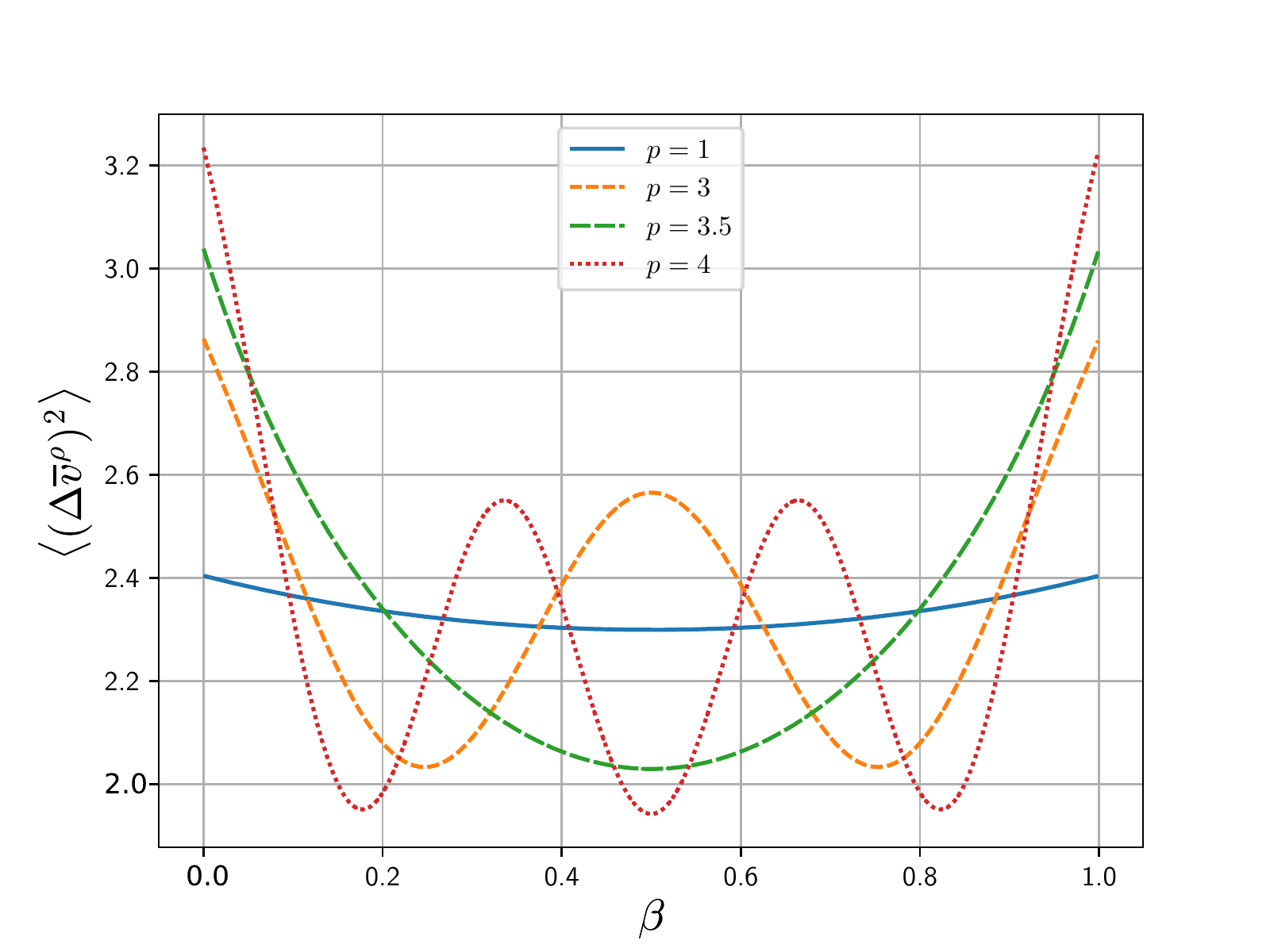}
\caption{The behavior of the dimensionless MSVD of the particle, $\langle(\Delta\overline{v}^{\rho})^{2}\rangle$, as a function of the dimensionless distance parameter $\chi$ for different combinations of the parameters $\beta$ and $p$ (left side), and  as a function of the quasiperiodic  parameter $\beta$ (right side). For the latter we have taken $\chi=1$.}\label{fig01}
\end{figure}

In Fig. \eqref{fig01} we plot, in the left side, the $\rho-$component of the dimensionless MSVD of the particle as function of $\chi$, for several different values of $\beta$ and $p$. In this case, for some combination of $\beta$ and $p$ we have plot regions where the magnitude of the velocity dispersion undergoes a decreasing or increasing, indicating nontrivial behavior. We note that, asymptotically, Eq. \eqref{eq307a} goes to a constant value, which corresponds to the configuration $\beta = 0$ and $p = 1$, associated with a periodic solution and no disclination. In fact, too far away from the defect, the influence of the disclination on the QBM of the particle is negligible. On the other hand, Fig. \eqref{fig01} also shows, in the right side, the plot of the $\rho-$component of the dimensionless MSVD of the particle as function of the quasiperiodic parameter $\beta$, taken $\chi=1$. The plot revels an oscillatory shape, which obviously is a consequence of the sinusoidal functions present in Eq. \eqref{eq307a}. In other words, this plot shows the highest or lowest values that Eq. \eqref{eq307a} can achieve for each value of $p$ and fixed distance $\chi$.

Another important feature associated with the $\rho-$component of the MSVD of the particle is its behavior near the disclination, $\rho=0$. In order to develop this analysis we have to investigate the behavior of the function $\mathcal{F}s_{y}(\chi)$, present in the integral definition of $\langle(\Delta\overline{v}^{\rho})^{2}\rangle$, Eq. \eqref{eq307a}. In fact, investigating the integrand in the limit $\chi\to 0$, the finiteness of the integral is  
dominated by large values of $y$, i.e., for $y\gg 1$. In this limit we can make use of the approximation $\cosh(y)\approx e^{y}/2$  and, as a consequence, we are able to show that the integral converge as long as $p$ and $\beta$ satisfy the restrictions $p\beta > 1$ and $p(1-\beta) > 1$. Note that for $\beta \in [0, 1/2)$ the first inequality includes the second, but if $\beta \in (1/2, 1)$ the opposite occurs, and only when $\beta = 1/2$ both inequalities are equivalents. The plots in the left side of Fig.  \eqref{fig01}, are in agreement with this analysis.

%
\subsection{MSVD in the $\phi$-direction}
%

\begin{figure}
\includegraphics[scale=0.5]{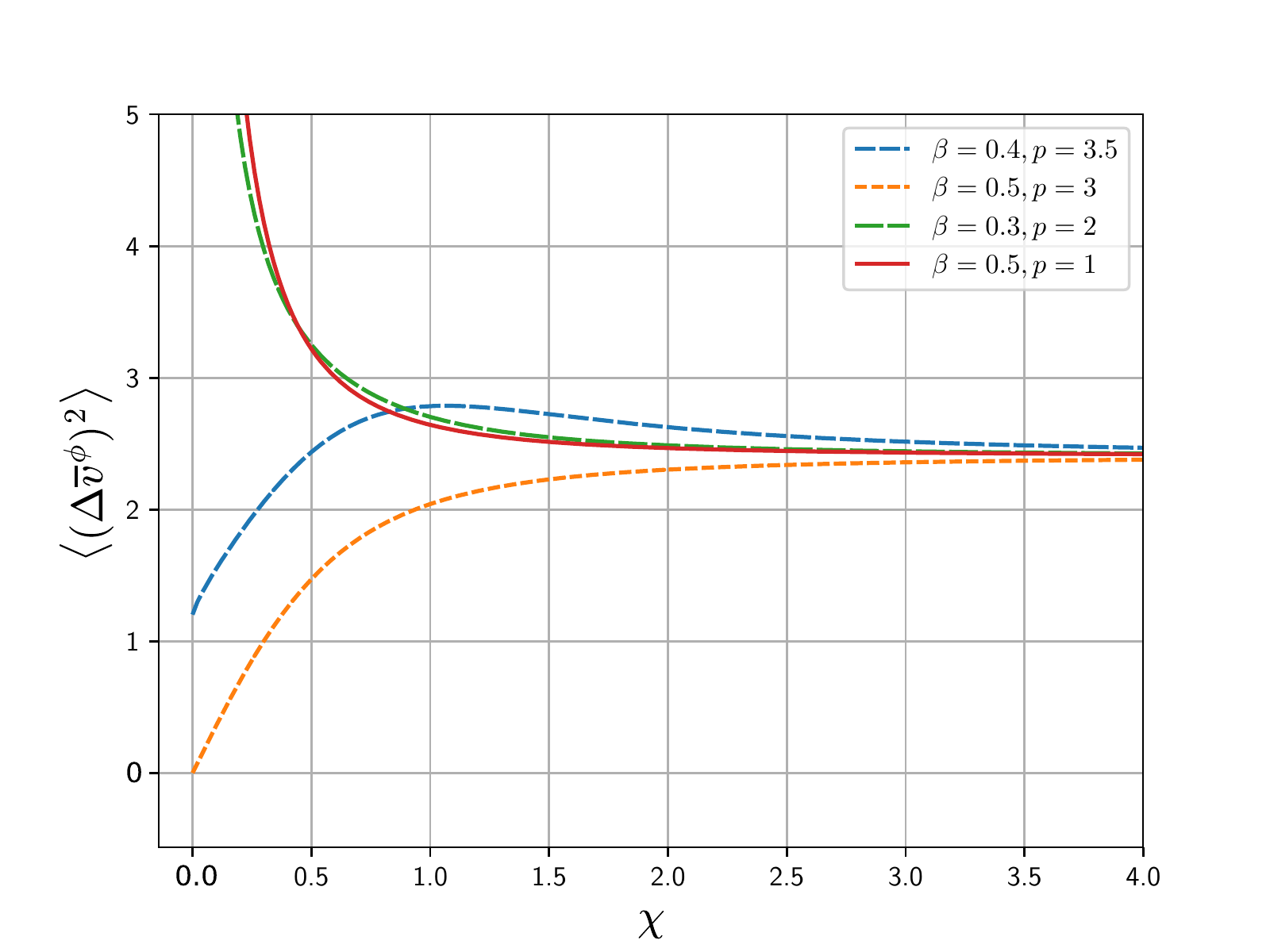}
\includegraphics[scale=0.5]{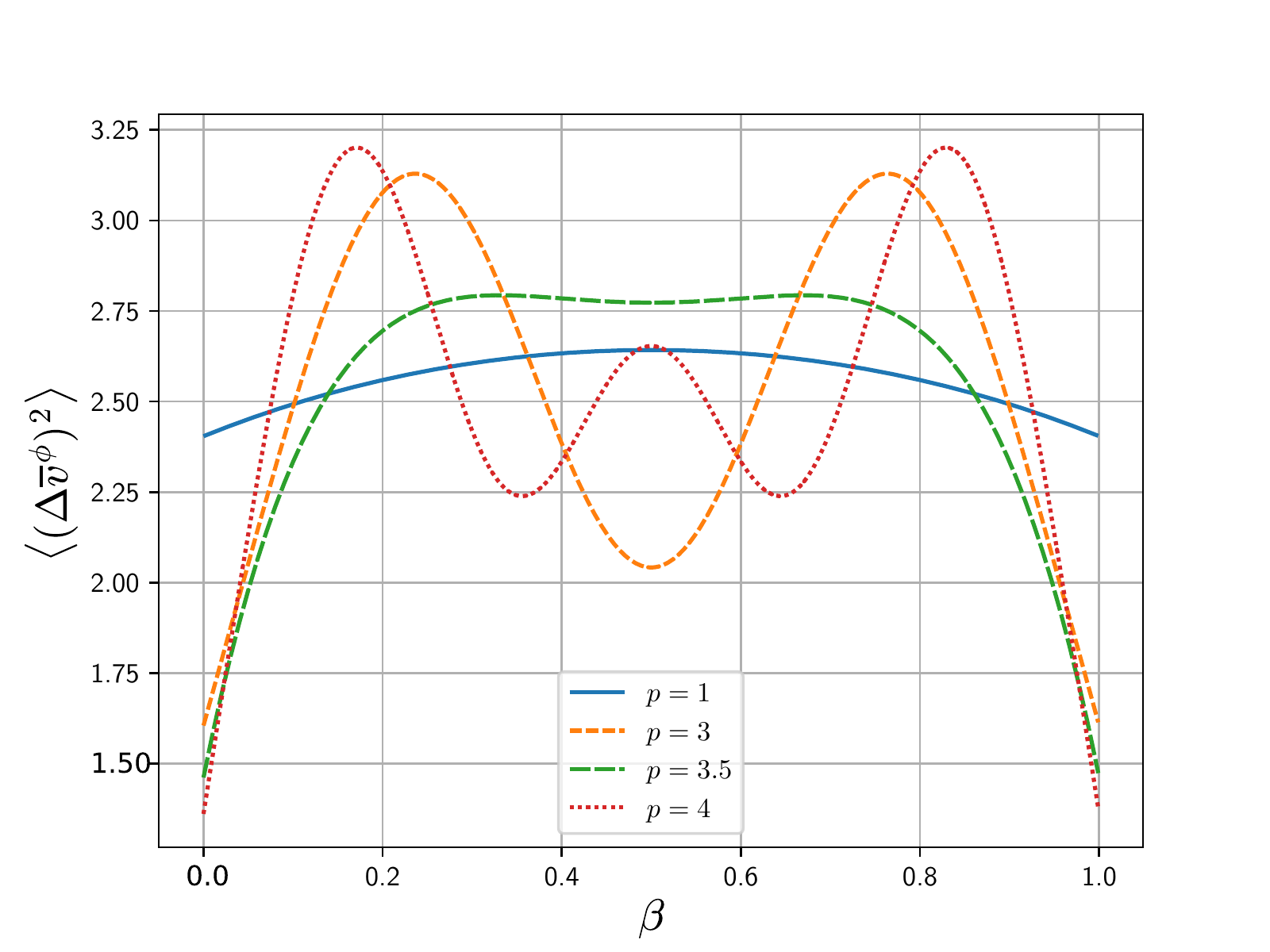}
\caption{The behavior of the dimensionless MSVD of the particle, $\langle (\Delta \overline{v}^{\phi})^{2} \rangle$, as a function of the dimensionless distance parameter $\chi$ (left side) and also as a function of the quasiperiodic  parameter $\beta$ (right side). For the latter we have taken $\chi=1$.}
\label{fig02}
\end{figure}

Now we turn to the calculation of the $\phi$-component of the MSVD of the particle. Thereby, in terms of the physical velocity $v^{\phi}=b_{\text{f}}^{-\frac{1}{4}}\rho u^{\phi}$, from \eqref{eq304}, we have
\begin{eqnarray}\label{eq308a}
\langle (\Delta v^{\phi})^{2} \rangle = \dfrac{q^{2}b_{\text{f}}^{\frac{1}{2}}}{m^2\rho\rho'}\int_{-\infty}^{\infty}d\eta_{2}b^{-\frac{1}{4}}(\eta_{2})\int_{-\infty}^{\infty}d\eta_{1}b^{-\frac{1}{4}}(\eta_{1})\partial_{\phi}\partial_{\phi'}\mathcal{W}(x,x')_{\text{d}},
\end{eqnarray}
where we have again extended the limits of integration from $-\infty$ to $\infty$. In order to solve the expression above we follow a similar procedure to that one used for the $\rho$-component, along with the method adopted in the Appendix A of Ref. \cite{mota2020induced}. Thus, in the coincidence limit, we obtain
\begin{eqnarray}\label{eq308b}
\langle (\Delta \overline{v}^{\phi})^{2} \rangle = 2\zeta(3)+2\sum_{m=1}^{\left[\frac{p}{2}\right]}\cos(2\pi m\beta)h_{m}(\chi) - \dfrac{p}{2\pi i}\int_{0}^{\infty}dy\mathcal{F}(\beta , p,y)h_{y}(\chi),
\end{eqnarray}
where the function $\mathcal{F}$ is given by Eq. \eqref{eq307b}, and the functions $h_{m}(\chi)$ and $h_{y}(\chi)$ are given in a compact form by
\begin{eqnarray}\label{eq308d}
h_{\mu}(\chi)=(1 - 2z_{\mu}^2)S_{2}(\mu,\chi) - \chi^{2}\left[1-(1 - 2z_{\mu}^2)^2\right]S_{3}(\mu,\chi)  \  ,
\end{eqnarray}
with $z_{\mu}=(z_m, z_y)=\left(\sin(m\pi/p),\cosh(y/2)\right)$ and the function $S_{\gamma}(\mu,\chi)$ has been defined in Eq. \eqref{eq307g}. Note that the MSVD of the particle in Eq. \eqref{eq308b} is in a dimensionless form and the constant responsible for that is the same as the one present in Eq. \eqref{eq307c}. In other words, in Eq. \eqref{eq307c} we must only replace $v^{\rho}$ with $v^{\phi}$ to obtain $\langle (\Delta \overline{v}^{\phi})^{2} \rangle$. The analysis about the finiteness of the  $\phi$-component of the MSVD of the particle at the disclination position, provides the same results as to the $\rho$-component. As consequence, the $\phi$-component of the MSVD in \eqref{eq308b} exhibits similar behaviors near $\rho=0$, as the one for the $\rho$-component, as we can see in the plots presented in Fig. \eqref{fig02}. This includes the discussion about the convergence or divergence, at $\chi=0$, of the velocity dispersion depending on the combination values for $p$ and $\beta$, which obey the restrictions $p\beta > 1$ and $p(1-\beta) > 1$ also applied here. So the analysis of these plots is essentially the same as that made below Eq. \eqref{eq307f} until the end of the sub-section.

\subsection{MSVD in the $z$-direction}

In terms of the physical velocity dispersion a long the $z$-direction, $v^{z}=b_{\text{f}}^{-\frac{1}{4}}u^{z}$, which is the parallel direction to the disclination, we can obtain the MSVD of the particle from Eq. \eqref{eq304}, resulting in
\begin{eqnarray}\label{eq309a}
\langle (\Delta v^{z})^{2} \rangle = \dfrac{q^{2}b_{\text{f}}^{\frac{1}{2}}}{m^2}\int_{-\infty}^{\infty}d\eta_{2}b^{-\frac{1}{4}}(\eta_{2})\int_{-\infty}^{\infty}d\eta_{1}b^{-\frac{1}{4}}(\eta_{1})\partial_{z}\partial_{z'}\mathcal{W}(x,x')_{\text{d}},
\end{eqnarray}
where we have extended the limits of integration to be from $-\infty$ to $+\infty$. To solve the integrals we follow the same steps as the ones in the calculation of the $\rho$-component. Thereby, in the coincidence limite $x'\rightarrow x$, we obtain 
\begin{eqnarray}\label{eq309b}
\langle (\Delta \overline{v}^{z})^{2} \rangle = 2\zeta(3)+2\sum_{m=1}^{\left[\frac{p}{2}\right]}\cos(2\pi m\beta)S_{2}(m,\chi) - \dfrac{p}{2\pi i}\int_{0}^{\infty}dy\mathcal{F}(\beta , p,y)S_{2}(y,\chi),
\end{eqnarray}
which is a dimensionless MSVD defined exactly in the same way as that in Eq. \eqref{eq307c}, of course, replacing $\rho$ by $z$ for the component. 
\begin{figure}
\includegraphics[scale=0.5]{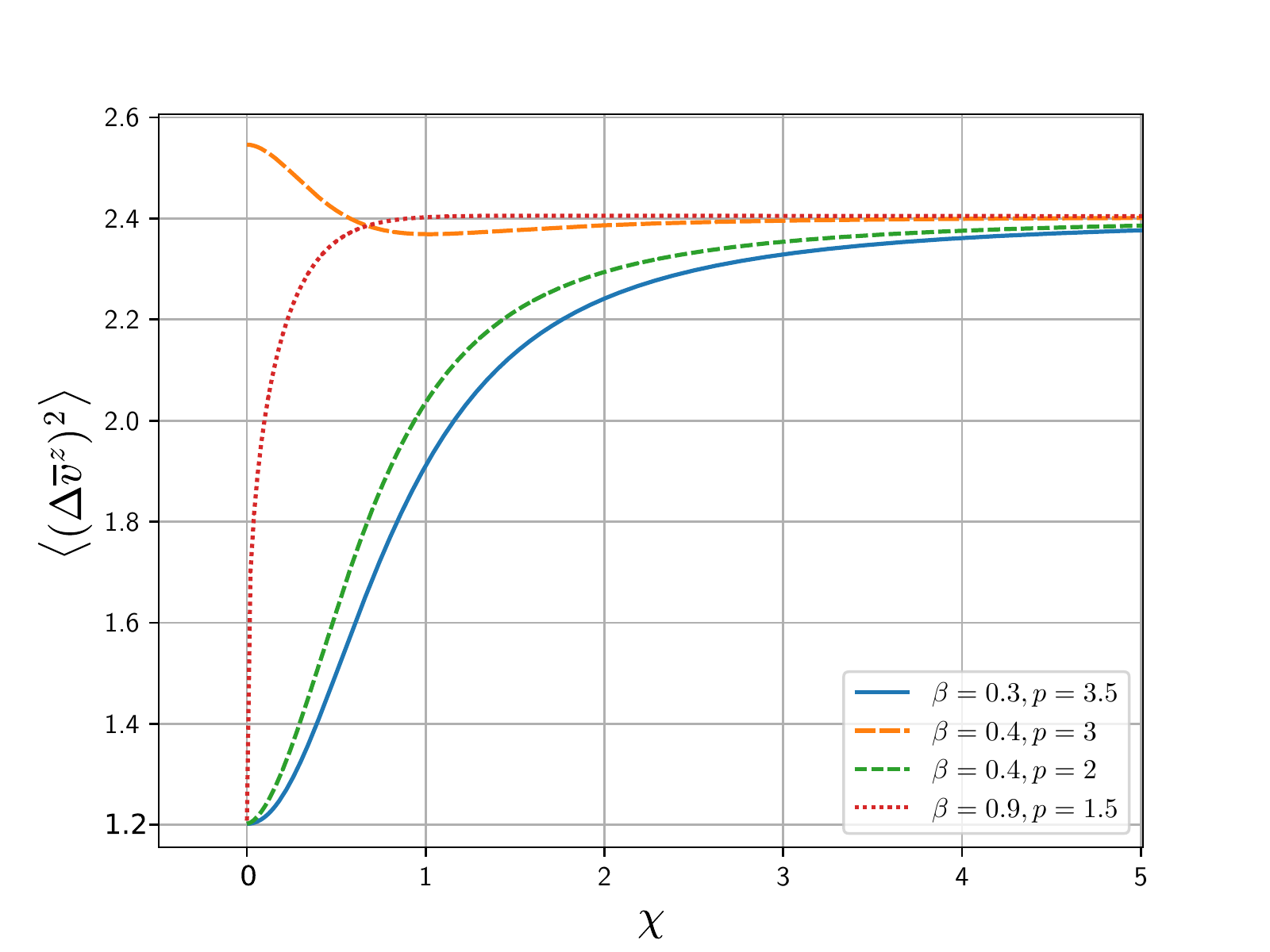}
\includegraphics[scale=0.5]{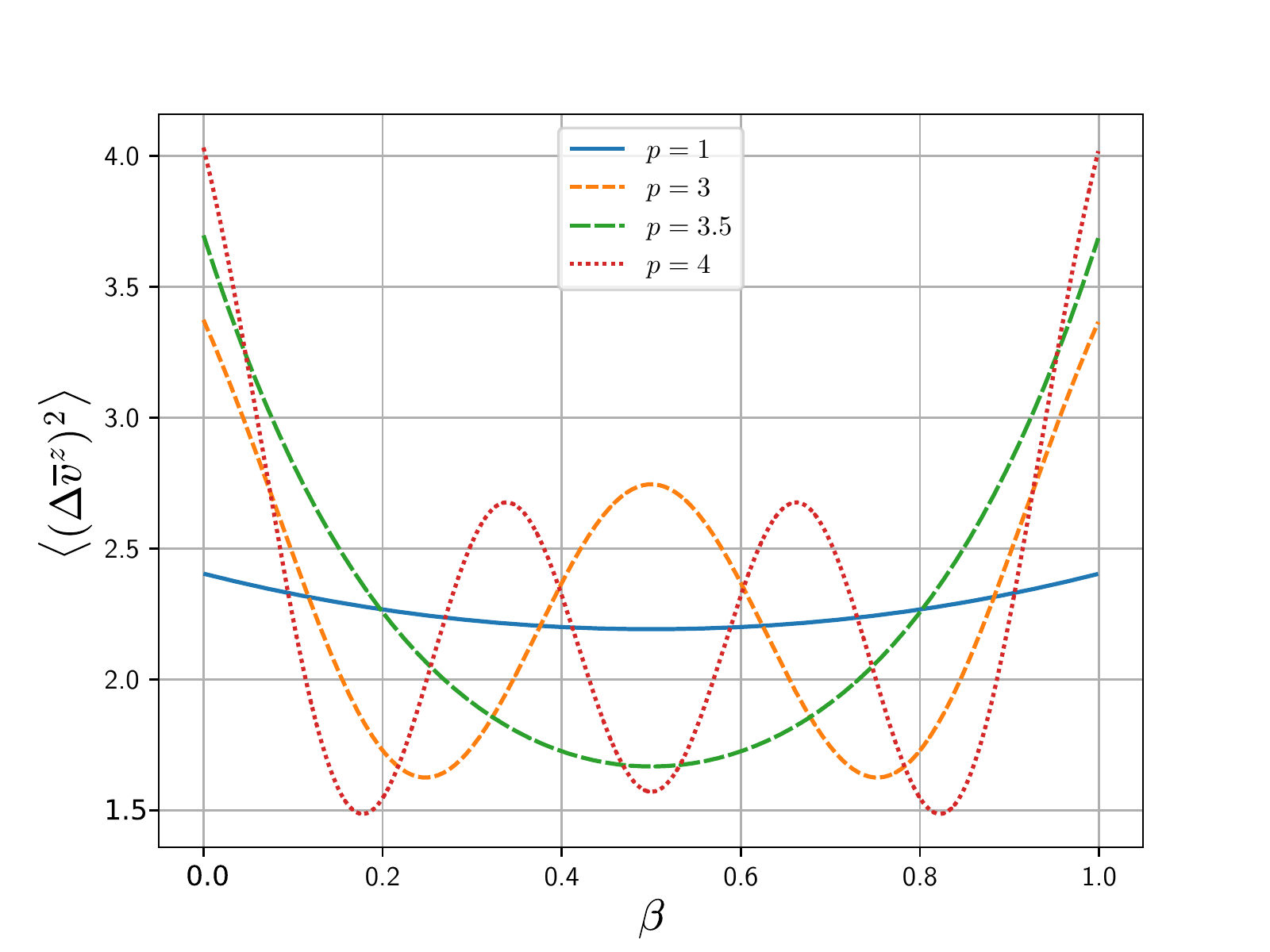}
\caption{Plot of the dimensionless velocity dispersion $\langle (\Delta \overline{v}^{z})^{2} \rangle$ as a function of the dimensionless distance parameter $\chi$ (left side) and quasiperiodicity parameter $\beta$ (right side). In the graph on the right side $\chi = 1$.}\label{fig03}
\end{figure}

In the left side of Fig. \eqref{fig03} we have plotted the MSVD, $\langle (\Delta \overline{v}^{z})^{2} \rangle$, as a function of $\chi$ for different values of $p$ and $\beta$, while in the right side we have plotted it in terms of the quasiperiodic parameter $\beta$, taken $\chi=1$. Differently from the other two components, the $z$-component of the MSVD of the particle in Eq. \eqref{eq309b} is always finite at $\chi=0$, for any values of $p$ and $\beta$ since the integrand depends only on the convergence of the function $\mathcal{F}$, which always happens. Besides that, for comparison effects, all the three components assume a constant value far way from the disclination, that is, $\chi\gg 1$. This value, as we can easily verify in the expressions for all three components is given by $2\zeta(3)$. 

\section{Conclusions}\label{sec07conclusion}

In this work we have studied the QBM of a point particle as a consequence of modifications on the quantum vacuum fluctuations of a massless real scalar field characterizing sound wave quantum excitations of a BEC. In this context, effectively, the condensate is described by the propagation of phonons in an analog FRW spacetime. In addition, we have also taken into consideration the presence of a disclination and a quasiperiodic condition on the angular variable that must be satisfied by the massless scalar field representing the phonons. Thus, in order to investigate the effects of this system on the motion of the point particle we have calculated its velocity dispersion.

The MSVDs of the scalar point particle that we have obtained are for a late time regime since we have extended the limits of integration of Eqs. \eqref{eq306}, \eqref{eq308a} and \eqref{eq309a} from $-\infty$ to $\infty$, resulting in a time independence of the corresponding expressions. The MSVDs present nontrivial behavior in all directions, but showed a global aspect, namely, a constant contribution given by $ 2\zeta(3)$ for distances far way from the disclination, i.e., $\chi\gg 1$. We have also verified that the expressions \eqref{eq307a} and \eqref{eq308b} for the velocity dispersion in the $\rho$ and $\phi$ directions may present either a divergent or convergent behavior at $\chi=0$, depending on the combination values of $p$ and $\beta$. In fact, a careful analysis have shown that in order to \eqref{eq307a} and \eqref{eq308b} be convergent the parameters $p$ and $\beta$ must obey the restrictions $p\beta>1$ and $p(1-\beta)>1$, otherwise they are divergent. This aspect is shown in Figs. \eqref{fig01} and \eqref{fig02}. In contrast, the expression \eqref{eq309b} for the $z$-component is always convergent and finite at $\chi=0$, for all values of $p$ and $\beta$, as shown in Fig. \eqref{fig03}. This occurs as a consequence of the integration in $y$ of the function $\mathcal{F}$ always be convergent at $\chi=0$, regardless the values of $p$ and $\beta$.

The constant behavior present for $\chi\gg 1$, for all three components of the MSVD of the particle, is actually the case in which there is no quasiperiodicity $(\beta=0)$ neither disclination $(p=1)$, providing a isotropic and homogeneous velocity dispersion. This constant result, $2\zeta(3)$, cannot be interpreted as corresponding to the Minkowski contribution, which is always divergent in the coincidence limit $x'\rightarrow x$. Note that this would be the case if  $b(t)=1$, corresponding to a non expanding condensate. In this sense, the element that produces the constant result is the choice for the scaling function $b(t)$ in Eq. \eqref{eq305}. Hence, the function $b(t)$ plays a similar role as the switching functions considered in Ref. \cite{camargo2018vacuum}. Therefore, the only divergencies arising here are related to the presence of the disclination, at $\chi=0$.

Finally, it is important to point out that, in the current study, both thermal and backreation effects were neglected. A more realistic treatment should be considered taken into consideration these effects. This will be presented elsewhere. 

{\acknowledgments}
E.J.B.F would like to thank the Brazilian agency CAPES for financial support. E.R.B.M is partially supported by CNPq under Grant no 301.783/2019-3. The author H.F.S.M. is supported by the Brazilian agency CNPq under Grants No. 305379/2017-8 and No. 311031/2020-0.


\end{document}